\documentclass[12pt] {article}
\usepackage{latexsym}
\usepackage{amsmath,amsfonts,amssymb,mathrsfs}
\usepackage{graphicx,epsfig}
\usepackage{here,cite}

\usepackage[top=2.5cm, left=2cm, right=2.6cm, totalheight = 23.5cm]{geometry}

\newcommand{\bea}{\begin{eqnarray}}
\newcommand{\eea}{\end{eqnarray}}
\newcommand{\be}{\begin{equation}}
\newcommand{\ee}{\end{equation}}

\newcommand{\ba}{\begin{align}}
\newcommand{\ea}{\end{align}}

\newcommand\T{\rule{0pt}{2.4ex}}        
\newcommand\BB{\rule[-0.0ex]{0pt}{0pt}}  

\begin{document}

\begin{titlepage}

\vspace*{1cm}%
\rightline{IFIC/07-48}%
\vspace{1cm}

\begin{center}

{\bf {\large Observational Evidence for Negative-Energy Dust \\
in Late-Times Cosmology} }\\
\vspace{1.2cm}%
{\bf Nikolaos~E.~Mavromatos$^{1}$ and Vasiliki~A.~Mitsou}$^{2}$ \\
\vspace{0.5cm}%
{\it $^1${King's College London, Physics Department, Theoretical Physics, Strand WC2R~2LS, UK} \\
\vspace{0.1cm}%
$^2${Instituto de F\'{i}sica Corpuscular (IFIC), CSIC -- Universitat de Val\`{e}ncia, \\
Edificio Institutos de Paterna, P.O.\ Box 22085, E-46071 Valencia, Spain} }\\

\vspace{1.2cm}%
{\bf Abstract}

\end{center}

We perform fits of unconventional dark energy models to the available data from
high-redshift supernovae, distant galaxies and baryon oscillations. The models
are based either on brane cosmologies or on Liouville strings in which a
relaxation dark energy is provided by a rolling dilaton field (Q-cosmology). An
interesting feature of such cosmologies is the possibility of effective
four-dimensional negative-energy dust and/or exotic scaling of dark matter. An
important constraint that can discriminate among models is the evolution of the
Hubble parameter as a function of the redshift, $H(z)$. We perform fits using a
unifying formula for the evolution of $H(z)$, applicable to different models.
We find evidence for a negative-energy dust at the current era, as well as for
exotic-scaling ($a^{-\delta}$) contributions to the energy density, with
$3.3\lesssim\delta\lesssim4.3$. The latter could be due to dark matter coupling
with the dilaton in Q-cosmology models, but it is also compatible with the
possibility of dark radiation from a brane Universe to the bulk in brane-world
scenarios, which could also encompass Q-cosmology models. The best-fit model
seems to include an $a^{-2}$-scaling contribution to the energy density of the
Universe, which is characteristic of the dilaton relaxation in Q-cosmology
models, not to be confused with the spatial curvature contribution of
conventional cosmology. We conclude that Q-cosmology fits the data equally well
with the $\Lambda$CDM model for a range of parameters that are in general
expected from theoretical considerations.

\vspace{2.5cm}

\leftline{July 31, 2007}

\end{titlepage}

\setcounter{page}{2}

\section{Introduction and summary}

There is a plethora of astrophysical evidence today, from supernovae
measurements~\cite{snIa,riess,SNLS,HST,essence}, the spectrum of fluctuations
in the Cosmic Microwave Background (CMB)~\cite{wmap}, baryon
oscillations~\cite{baryon} and other astrophysical data, indicating that the
expansion of the Universe is currently accelerating. The energy budget of the
Universe seems to be dominated at the present epoch by a mysterious dark energy
component, but the precise nature of this energy is still unknown. Many
theoretical models provide possible explanations for the dark energy, ranging
from a cosmological constant~\cite{carroll} to super-horizon
perturbations~\cite{riotto} and time-varying quintessence
scenarios~\cite{steinhardt}, in which the dark energy is due to a smoothly
varying (scalar) field which dominates cosmology in the present era.

The current astrophysical data are capable of placing severe constraints on the
nature of the dark energy, whose equation of state may be determined by means
of an appropriate global fit. Most of the analyses so far are based on
effective four-dimensional Robertson-Walker Universes, which satisfy on-shell
dynamical equations of motion of the Einstein-Friedman form. Even in modern
approaches to brane cosmology~\cite{brane}, which are described by equations
that deviate during early eras of the Universe from the standard Friedman
equation (which is linear in the energy density), the underlying dynamics is
assumed to be of classical equilibrium (on-shell) nature, in the sense that it
satisfies a set of equations of motion derived from the appropriate
minimisation of an effective space-time Lagrangian.

However, cosmology may not be an entirely classical equilibrium situation.  The
initial Big Bang or other catastrophic cosmic event, which led to the initial
rapid expansion of the Universe, may have caused a significant departure from
classical equilibrium dynamics in the early Universe, whose signatures may
still be present at later epochs including the present era. In this context,
there has been proposed a specific model for the cosmological dark energy,
being associated with a rolling dilaton field that is a remnant of this
non-equilibrium phase, described by a generic non-critical string
theory~\cite{emnw,diamandis,diamandis2,lmn}. We call this scenario
`Q-cosmology'.

Since such a non-equilibrium, non-classical theory is \emph{not} described by
the equations of motion derived by extremising an effective space-time
Lagrangian, one must use a more general formalism to make predictions that can
be confronted with the current data. The approach we favour is formulated in
the context of string/brane theory~\cite{gsw,polchinski}, the best candidate
theory of quantum gravity to date. Our approach is based on non-critical
(Liouville) strings~\cite{aben,ddk,emn}, which offer a mathematically
consistent way of incorporating time-dependent backgrounds in string theory.

The basic idea behind such non-critical Liouville strings is the following.
Usually, in string perturbation theory, the target space dynamics is obtained
from a stringy $\sigma$-model~\cite{gsw} that describes the propagation of
strings in classical target-space background fields, including the space-time
metric itself.  Consistency of the theory requires conformal invariance on the
world sheet, in which case the target-space physics is independent of the scale
characterising the underlying two-dimensional dynamics. These conformal
invariance conditions lead to a set of target-space equations for the various
background fields, which correspond to the Einstein/matter equations derived
from an appropriate low-energy effective action that is invariant under general
coordinate transformations. Unfortunately, one cannot incorporate in this way
time-dependent cosmological backgrounds in string theory, since ---to low
orders in a perturbative expansion in the Regge slope $\alpha '$--- the
conformal invariance condition for the metric field would require a Ricci-flat
target-space manifold, whereas a cosmological background necessarily has a
non-vanishing Ricci tensor.

To remedy this defect, and thus be able to describe a time-dependent
cosmological background in string theory, the authors of Ref.~\cite{aben}
suggested that a non-trivial r\^ole should be played by a time-dependent
dilaton background.  This approach leads to strings living in numbers of
dimensions different from the customary critical number, and was in fact the
first physical application of non-critical strings~\cite{ddk}. The approach of
Ref.~\cite{aben} was subsequently extended~\cite{emn,emnw, diamandis,
diamandis2,lmn} to incorporate off-shell quantum effects and non-conformal
string backgrounds describing other non-equilibrium cosmological situations,
including catastrophic cosmic events, such as the collision of two brane
worlds.

In a recent work~\cite{EMMN}, we have presented preliminary constraints on
Q-cosmology by means of supernova data. An important result of our analysis was
the unavoidable presence of negative-dust-like contributions to the energy
budget of the Universe. Such negative-energy dust terms arise naturally in
Q-cosmology~\cite{EMMN}, as a result of the existence of non-equilibrium dark
energy contributions. However, a similar feature may also characterise certain
brane cosmologies as a result of effective four-dimensional Kaluza-Klein (KK)
graviton mode contributions to the energy density of the brane~\cite{kaluza}.
Such brane models are also characterised by dark radiation $a^{-4}$ terms, as a
result of non-trivial gravitational bulk dynamics~\cite{brane}. As discussed in
Ref.~\cite{EMMN}, an important distinguishing feature of Q-cosmology is an
$a^{-2}$-scaling contribution to the dark energy as a result of dilaton
relaxation terms~\cite{emnw}. Such a term is distinct from the spatial
curvature contribution as we shall review below. The astrophysical data
analysis of Ref.~\cite{EMMN} yields a non-zero value of such dark energy
dilaton terms.

It is the purpose of the current work to update and complete such analyses by
taking into account the very recent supernova data~\cite{HST,essence}, as well
as data from (luminous red) galaxies~\cite{baryon,Hz,SDSS} on the evolution of
the Hubble parameter as a function of the redshift, which, as we shall see,
turns out to be an important constraint for model building. In the same manner
as in the analysis of Ref.~\cite{EMMN}, we also perform a comparison of the
results of the fit for the Q-cosmology model with those of the conventional
$\Lambda$CDM model and a rival model with super-horizon
perturbations~\cite{riotto} superposed on an underlying
Einstein-Friedman-Robertson-Walker Universe. We find that the current data are
consistent with {\it both} the Q-cosmology and $\Lambda$CDM models. On the
other hand, for the super-horizon model there appears a $2\sigma$
incompatibility between the best-fit values for the $H(z)$ galactic data and
the supernova analysis. As in Ref.~\cite{EMMN}, the current analysis indicates
the existence of negative-energy dust and non-trivial $a^{-2}$ dark-energy
contribution attributed to the dilaton relaxation terms. Our results in this
work should be considered as complementary to other similar studies of
unconventional cosmologies that exist in the current literature~\cite{studies}.

The structure of the article is as follows. In Sec.~\ref{sc:qcosmo}, the most
crucial phenomenological aspects of brane and Q-cosmology models are outlined,
and a generic parametrization for the expansion rate of the Universe, to be
used in the experimental fits, is developed, which is capable of capturing, in
a single formula, the most important features of the various models, namely
negative-energy dust, exotic scaling components to the energy density and
dilaton dark energy relaxation terms. The cosmological data analyses are
discussed in Sec.~\ref{sc:analysis} for each of the three observational sources
under study, namely the supernovae, the $H(z)$ measurement from distant
galaxies observations and the baryon acoustic oscillations. In
Sec.~\ref{sc:discussion}, the astrophysical constraints on the cosmological
models are combined, leading to the determination of the model parameters, and
implications on possible unconventional cosmological scenarios are discussed.
Finally, conclusions and outlook are presented in Sec.~\ref{sc:conclu}.

\section{Theoretical models}\label{sc:qcosmo}

\subsection{Brane Universes}

We commence our discussion by summarising the basic features of brane
cosmology~\cite{brane}. According to this picture, matter fields are confined
on three-brane worlds, while fields from the gravitational multiplet (dilaton,
graviton, \emph{etc.}) are allowed to propagate in the bulk. For definiteness,
in this work we consider five-dimensional models. In such a case, the analogue
of the effective four-dimensional Friedman equation on the brane
reads~\cite{brane}:
\begin{equation}\label{eq:brane}
    H^2 = \frac{8\pi G}{3} \rho_{\rm M} \left( 1+\frac{\rho_{\rm M}}{2\sigma} \right)
    + \frac{\Lambda_4}{3} + \frac{\mu}{a^4},
\end{equation}
where $\rho_{\rm M}$ is the matter density on the brane, and we have identified
the brane tension $\sigma$ with the Newton's constant $8\pi G/3 = \sigma/18$
and the four-dimensional cosmological constant $\Lambda_4$ is related to the
AdS-bulk (negative) cosmological constant $\Lambda_5$ as follows:
\begin{equation}\label{eq:lambda}
    \frac{\Lambda_4}{3} = \frac{\sigma^2}{36} + \frac{\Lambda_5}{6}.
\end{equation}

For our purposes, we shall assume a fine tuning such that $\Lambda_4=0$. The
term $\mu/a^4$ in~(\ref{eq:brane}) is known as dark radiation, and stems from
the non-trivial bulk dynamics and energy conservation. Quadratic terms in
$\rho_{\rm M}$ do not play a crucial r\^ole in late times, and hence can be
ignored when discussing phenomenology for redshifts below three, which will be
of interest to us in this work.

An interesting feature of brane models has been pointed out in
Ref.~\cite{kaluza} and concerns the contributions to the effective
four-dimensional energy density on the brane world by KK graviton modes. Such
modes appear as massive gravitons on the brane with masses $m>3H/2$, where $H$
represents the expansion rate of a de Sitter brane world we assume here for
concreteness. The analysis of Ref.~\cite{kaluza} has shown that a single very
massive KK mode, corresponding to a particle with high momentum along the bulk
direction, exerts pressure on the brane pushing it outwards. As a result, the
energy in the bulk decreases leading to a decrease in the dark energy term, and
therefore an effective negative contribution to the dark radiation term. In
this way, the KK mode behaves like a negative-energy dust. It should be noted
at this stage that this analysis is still incomplete and cannot be trivially
extended to realistic cosmological models, where one should integrate over the
entire spectrum of the KK modes. Nonetheless, for our phenomenological purposes
in this work we assume that KK modes behave like negative-energy dust. It
should be noted that there is no conflict with positive-energy theorems, since
the bulk energy density contribution of the KK modes is positive~\cite{kaluza}.

\subsection{Q-cosmology}

Next we discuss briefly Q-cosmologies, which is our second class of models to
be constrained by the data. Details on the dynamical equations describing
Q-cosmological models with matter and dark energy contributions have been
reviewed in Ref.~\cite{EMMN}, where we refer the interested reader. In the
analysis that follows, we will only outline the basic predictions for the
Hubble rate $H(z)$, which will be used in our fits.

For completeness, however, we stress again  that the most important feature of
Q-cosmology is the {\it off-shellness} of the appropriate equations describing
the dynamics of the fields in the gravitational string multiplet (graviton and
dilatons). The target-space effective action variations with respect to the
gravitational field are non zero, and the respective values are determined by
the requirement that the Liouville mode restores the world-sheet conformal
invariance of the non-critical string. The identification of the Liouville mode
with the target time leads to a set of dynamical equations and
constraints~\cite{emnw,diamandis,diamandis2}, whose consistent solution
determines the Liouville string Universe, under the standard assumptions of
homogeneity and isotropy.

There are several versions of the Liouville theory, which in general lead to
different predictions. There may be models in which all matter and
gravitational fields are non conformal on the world sheet, or others in which
only the gravitational multiplet deviate from criticality. For our purposes in
this work and in Ref.~\cite{EMMN}, we restrict ourselves to the latter
category. After the identification of the Liouville mode with the target time,
which can be explained dynamically in such models~\cite{gravanis}, the
pertinent equations for gravitons, for instance near a fixed point in theory
space (and thus for sufficiently long cosmic times), read:
\begin{eqnarray}
0 \neq \frac{\delta S^{G,{\rm matter}} }{\delta g^{\mu\nu}} = {\ddot
g}_{\mu\nu} + Q(t){\dot g_{\mu\nu}} + {\cal O}(g^2)~, \label{eq:liouveqs}
\end{eqnarray}
where $S^{G,{\rm matter}}$ denotes the gravitational and matter effective
low-energy action in the target space and the dot is a Liouville world-sheet
zero-mode derivative, which at the very end is identified (dynamically) with
the cosmic time. The central charge deficit $Q^2(t)$ describes the microscopic
non-critical model, and is expressed as a functional of graviton $g_{\mu\nu}$
and dilaton $\Phi$ fields of the gravitational string multiplet. To lowest
order in the Regge slope $\alpha '$, it is given by:
\begin{equation}
0 < Q^2 (t) = 3g^{00}\left({\ddot \varphi} - ({\dot \varphi})^2 + \dots~\right), \quad
\varphi \equiv 2\Phi - {\rm log}\sqrt{g}~.
\label{qdef}
\end{equation}
We note at this stage that the positive definiteness of $Q^2(t)$ (supercritical
string~\cite{aben}) is a model-dependent feature, but it is {\it essential} in
yielding a time-like Liouville mode, thus allowing its interpretation as a
target time field. Such supercritical strings may describe, for instance, the
stringy excitations on the three brane describing the observable Universe in a
colliding-brane-world scenario~\cite{emnw}, with the collision providing the
main cause for departure from criticality.

The off-shell terms of the Eqs.~(\ref{eq:liouveqs}) imply highly non trivial
modifications of the analogue of the (effective, four-dimensional) Friedman
equation for the Q-cosmology, which now includes, apart from the standard
matter and dark energy contributions denoted collectively by $\rho$, also the
above-mentioned Liouville-string {\it off-shell} corrections, $\Delta \rho$,
which are not positive definite in general (without affecting, however, the
overall positive-energy conditions of this non-equilibrium theory):
\begin{equation}
H^2(z) = \frac{8\pi G_N}{3}\rho + \Delta \rho
\label{friedmann}
\end{equation}
The appropriate parametrisation for $H(z)$ in the Q-cosmology framework at late
eras, such as the ones pertinent to the supernova and other data ($0<z<2$),
where some analytic approximations are allowed~\cite{EMMN}, reads:
\begin{equation}\label{eq:formulafit}
H(z) = H_0 \left( {\Omega }_3 (1 + z)^3 + {\Omega }_{\delta} (1 + z)^\delta +
{\Omega}_2(1 + z)^2 \right)^{1/2},\quad{\Omega }_3 + {\Omega}_{\delta} +
{\Omega}_2 = 1,
\end{equation}
with the densities $\Omega_{2,3,\delta}$ corresponding to present-day values
($z = 0$). It is important to remark that, as discussed in detail in
Ref.~\cite{EMMN}, the densities $\Omega_i$ involve {\it mixed} contributions
from {\it both}, {\it off-shell} dark energy and (dark) matter. In this
respect, $\Omega_2$ denotes the dilaton dark energy contribution at late eras,
not to be confused with spatial curvature contributions. Similarly, $\Omega_3$
does not denote matter dust, but it includes (negative) dark energy
contributions scaling with the redshift like dust. This is a result of the
coupling of the dilaton as well as higher string loop corrections, which,
depending on the model, could be significant at late epochs of the Universe. In
general, there is no conflict with positive energy theorems, in a similar
spirit to negative-energy dust contributions in brane models, as mentioned
previously, which can be due to appropriately compactified KK graviton
modes~\cite{kaluza}. We must note at this point that Q-cosmologies can indeed
be accommodated in brane-world scenarios. This may occur, for instance, in
colliding branes, where the initial brane collision constitutes an
Early-Universe cosmically catastrophic event, responsible for the departure of
the string theory on the brane from criticality~\cite{emnw}. In such brany
Q-cosmologies, KK graviton modes appear naturally, contributing negative-energy
dust components to the effective four-dimensional energy density on the brane.
Due to the dilaton dark-energy-relaxation (quintessence) terms, however, in
such models, one may also obtain in general additional scaling contributions of
the type $a^{-2}$ at late eras~\cite{emnw,EMMN}, which are absent in
conventional brane models with constant dilatons (c.f.\ (\ref{eq:brane})).

Moreover, in Q-cosmologies there are dark matter contributions with exotic
scaling, denoted by $\Omega_\delta$, where the exponent $\delta$ is
theoretically close to, but not quite, four; such contributions are due to the
coupling of dark matter species with the (non-constant) dilaton quintessence
field. We stress once more that in all the above terms, the off-shell Liouville
corrections play an important r\^ole~\cite{diamandis2,lmn}.

A complete analysis of the non-critical-string and dilaton effects, which turn
out to be important in the present era after the inclusion of matter, requires
a numerical treatment~\cite{diamandis2,lmn}. The energy budget of the
Q-cosmology Universe is at present largely phenomenological, since the
theoretical analysis, involving string loops, is not sufficiently developed to
allow for concrete predictions insofar as the values of $\Omega_i$ and $\delta$
are concerned. For instance, in Ref.~\cite{lmn} it was assumed, rather {\it ad
hoc}, that there is no exotic matter present today ($z = 0$), although this is
not true in the past. An important constraint is, of course, nucleosynthesis,
which in the analysis of Ref.~\cite{lmn} has been left essentially undisturbed,
occurring at MeV scales.

In general, in our analysis the three parameters to be determined by the fits
are $\Omega_3$, $\Omega_\delta$ and $\delta$. As we shall see, the data seem to
point towards the fact that the amount of exotic-scaling dark matter today is
non zero, in contrast to the assumption of Ref.~\cite{lmn}. This, in turn,
implies the need for revisiting the phenomenological constraints on
supersymmetric particle physics models at future colliders, such as the LHC,
derived in Ref.~\cite{lmn} within the Q-cosmology framework. We shall do so in
a future publication.

\subsection{Formulae for data analysis}

In our phenomenological fitting procedure, $\delta$ is left as a free
parameter, whose best fit value, however, is found, as we shall discuss below,
close to four, thus in good agreement with theoretical
expectations~\cite{diamandis,diamandis2,lmn}. It is important to notice that
such exotic matter scaling has important consequences on the relaxation of some
of the stringent constraints imposed on supersymmetric models of particle
physics from thermal dark matter analyses~\cite{lmn} in standard
Friedman-Robertson-Walker cosmologies. On the other hand, a value $\delta=4$
may correspond to dark radiation in brane models~\cite{kaluza}. In this sense,
our parametrisation~(\ref{eq:formulafit}) may be considered as a unifying
formula, which can be used for fitting various Q-cosmologies and brane models:
\begin{equation}\label{eq:formulaforfit}
\boxed{ \,H(z) = H_0 \left( {\Omega }_3 (1 + z)^3 + {\Omega }_{\delta} (1 +
z)^\delta + {\Omega}_2(1 + z)^2 \right)^{1/2} }
\end{equation}
with ${\Omega }_3 +{\Omega}_{\delta} + {\Omega}_2 = 1$ and ${\Omega }_3$,
${\Omega}_{\delta}$, $\delta$ as free parameters.

Before closing this section, we mention that we shall compare our results with
two other cosmological models: a $\Lambda$CDM cold dark matter model with a
cosmological constant~\cite{carroll}, where the Hubble parameter is expressed
as:
\begin{equation}
H(z) = H_0 \left(\Omega_{\rm M} ( 1 + z)^3 + \Omega_\Lambda ( 1 + z )^{3(1 +
w_0)}\right)^{1/2},\label{eq:lcdm}
\end{equation}
and the super-horizon model, in which the Universe is assumed to be filled with
non-relativistic matter only~\cite{riotto} and there is no dark energy of any
sort:
\begin{equation}
H(z) = \frac{{\overline H}_0}{1 - \Psi_{\ell 0}}\left(a^{-3/2} -
a^{-1/2}\Psi_{\ell 0}\right), \label{eq:shcdm}
\end{equation}
with $\Psi(\vec{x},t)$ the gravitational potential, $\Psi_{\ell 0}$ a free
parameter, $1 + z = a^{-1}(t) e^{(a(t)-1)\Psi_{\ell 0}}$ and
$a(t)e^{-\Psi_{\ell}(t)+\Psi_{\ell 0}}$ the scale factor of the
Robertson-Walker Universe.

\section{Data analysis}\label{sc:analysis}

Observational data from three different astrophysical sources ---namely
supernovae, distant galaxies and baryon acoustic oscillations--- are analysed
in order to set constraints on the aforementioned cosmological models. The data
fits and all analyses involved are performed within the ROOT~\cite{root}
analysis framework and the Minuit~\cite{minuit} minimisation and error
computation code.

In what follows, we shall use `generically' the terminology `Q-cosmology'
whenever we use formula~(\ref{eq:formulaforfit}) as a fitting function to refer
to both brane models with constant dilatons~(\ref{eq:brane}) and
dilaton-quintessence Q-cosmology models. The reader should bear in mind that
the distinguishing features of dilaton-quintessence Q-cosmology models, as
compared to brane models with constant dilatons, are the $a^{-2}$-scaling
contributions to the energy density of the Universe, as well as the exotic dark
matter scaling $a^{-\delta}$, with $\delta\neq4$.

Some important remarks are now in order. In the discussion of $\Lambda$CDM
model, we have allowed, for the sake of completeness, for non-zero spatial
curvature contributions. On the other hand, for simplicity, when we discuss
Q-cosmology and super-horizon models, we assumed a spatially flat Universe.
This is reflected in the relevant relations between observables and the Hubble
parameter to be discussed in the following sections. This should make even
clearer the distinct r\^ole played by the above mentioned dilaton dark energy
$a^{-2}$ contributions to $H(z)$, as compared to spatial curvature terms of
conventional cosmology.

\subsection{Supernovae}

We analyse recent type-Ia supernovae (SN) data released by the Hubble Space
Telescope (HST)~\cite{HST} and the ESSENCE collaboration~\cite{essence}. Among
16~newly discovered high-redshift SNe~\cite{HST}, the so-called `gold' dataset
(\emph{Riess07:gold}) includes SNe from other sets: 14~SNe discovered earlier
by HST~\cite{riess}, 47~SNe reported by SNLS~\cite{SNLS} and 105~SNe detected
by ground-based discoveries, amounting a total of 182 data points. An
additional set of 77~SNe tagged as `silver' due to lower quality of photometric
and spectroscopic record is also listed, constituting the
\emph{Riess07:gold+silver} dataset. The ESSENCE dataset (\emph{WV07}), on the
other hand, consists out of 60~SNe of $0.015<z<1.02$ discovered by
ESSENCE~\cite{essence}, 57~high-$z$ SNe discovered during the first year of
SNLS~\cite{SNLS} and 45~nearby SNe. Besides these SN sets, a compilation of the
aforementioned data (\emph{WV07+HST})~\cite{davis}, normalised in order to
account for the different light-curve-fitters employed, is eventually analysed.
The latter sample, listing 192~SNe, includes all high-redshift supernovae
($z>1$) observed so far.

Supernova data are given in terms of the distance modulus $\mu = 5\log d_L +
25$, where the luminosity distance $d_L$ (in megaparsecs) for a spatially flat
universe is related to the redshift $z$ via the Hubble rate $H(z)$:
\begin{equation}\label{luminositydistanceredshift}
d_L(z) = c(1 + z) \int_0^z \frac{dz'}{H(z')},
\end{equation}
whereas in the general case of a non-zero spatial curvature contribution
$\Omega_k$ the luminosity distance is given by~\cite{riess}:
\begin{multline}
\qquad\qquad\qquad d_L(z) = \frac{c(1+z)}{\sqrt{|\Omega_k|}} \times \begin{cases}%
 \sin\left( \sqrt{|\Omega_k|}\int_0^z \frac{dz'}{H(z')}\right), & \mbox{for a closed Universe,} \\
 \sqrt{|\Omega_k|}\int_0^z \frac{dz'}{H(z')}, & \mbox{for a flat Universe,} \\
 \sinh\left( \sqrt{|\Omega_k|}\int_0^z \frac{dz'}{H(z')}\right), & \mbox{for an open Universe.}
\end{cases}\label{luminositydistanceredshift_curv}
\end{multline}
We note that this observable depends on the expansion history of the Universe
from $z$ to the present epoch, and recall that, although most of the available
supernovae have $z < 1$, there is a handful with values of redshift up to
$z\simeq1.8$. For illustration purposes, henceforth both data and predictions
of cosmological models will be expressed as residuals, $\Delta\mu$, from the
empty-Universe prediction (Milne's model, $\Omega_{\rm M}=0$). The analysis
involves minimisation of the standard $\chi^2$ function with respect to the
cosmological model parameters, as they are introduced via the Hubble expansion
rate in Eqs.~(\ref{eq:formulaforfit}),~(\ref{eq:lcdm}) and~(\ref{eq:shcdm}),
for the Q-cosmology, $\Lambda$CDM and super-horizon models, respectively.

\begin{figure}[htb]
\begin{center}
\includegraphics[width=0.7\linewidth]{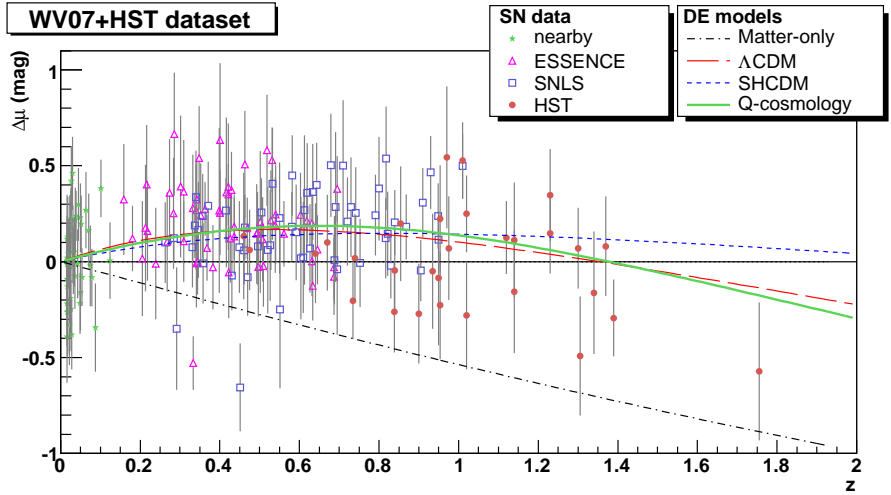}
\end{center}
\caption{\it Residual magnitude of the WV07+HST dataset supernovae~\cite{davis}
versus the redshift. Cosmological model predictions are drawn: (i) Milne's
Universe (black solid line); (ii) Matter-only Universe, $\Omega_{\rm M}=1$
(black dashed-dotted line); (iii) $\Lambda$CDM model for $(\Omega_{\rm M},
\Omega_\Lambda) = (0.33,0.85)$ (red long-dashed line); (iv) super-horizon model
for $\Psi_{\ell0} = -0.90$ (blue short-dashed line); and (v) Q-cosmology for
$\delta=4$, $\Omega_3=-2.8$ and $\Omega_\delta=0.86$ (green thick solid line).
} \label{fig:SNres_Davis07}
\end{figure}

The WV07+HST dataset, which amounts to a sample of 192~supernovae in total, is
shown in Fig.~\ref{fig:SNres_Davis07} with the respective measurement errors.
This sample combines all observed high-redshift supernovae, most of them
detected by the HST. The predictions of the cosmological models under study are
also displayed for the best-fit parameter values.

\begin{figure}[htb]
\begin{center}
\includegraphics[width=0.45\linewidth]{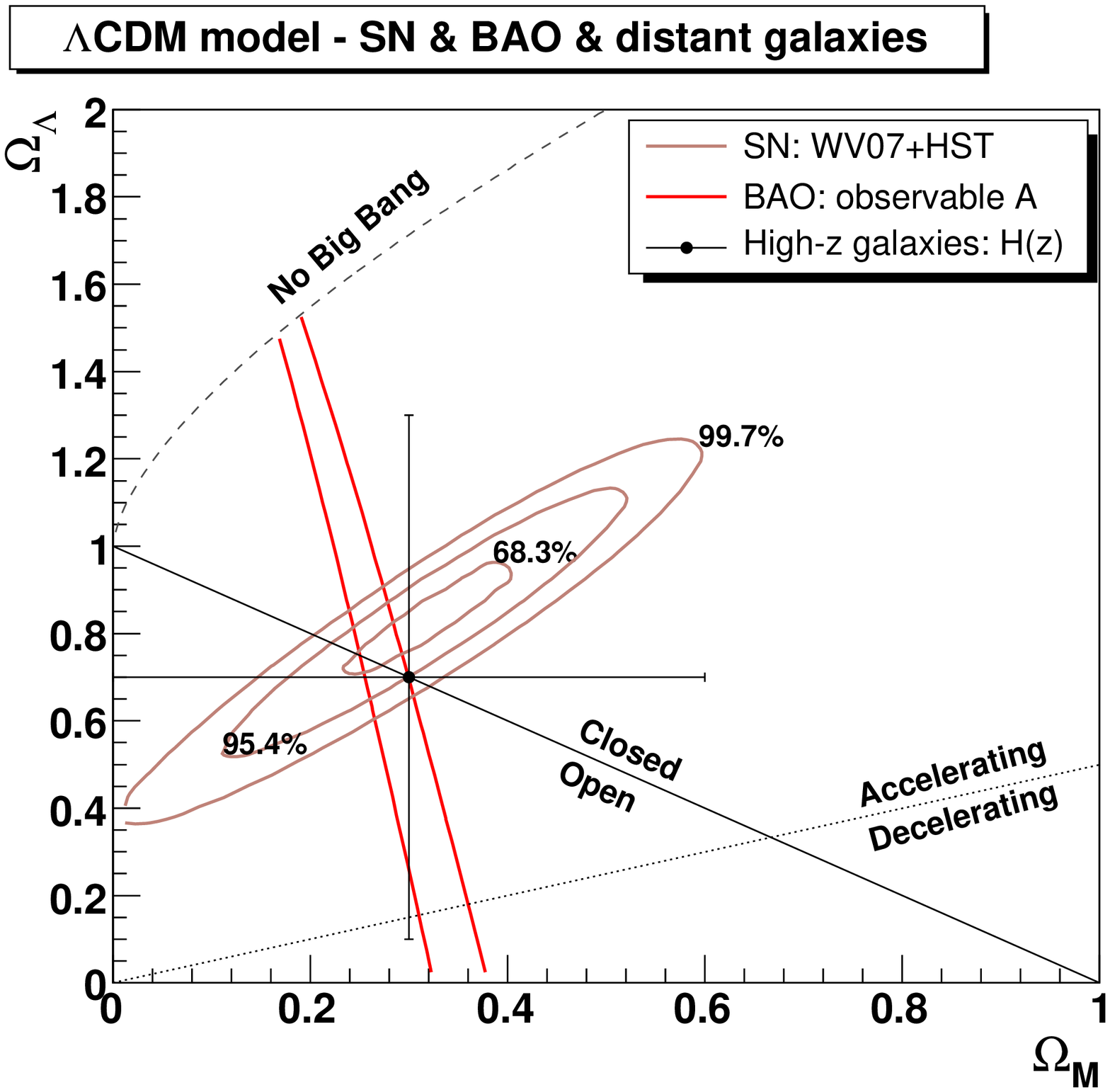}
\end{center}
\caption{\it Confidence-level contours for the $\Lambda$CDM model for the
WV07+HST sample of 192~supernovae~\cite{davis}. The 68.3\% C.L.\ region
extracted from the baryon-oscillations observable $A$~\cite{baryon} and the
result of the fit to the $H(z)$ measurement~\cite{Hz} from distant galaxies is
also superposed.} \label{fig:SN_BAO_Hz_LCDM_cont}
\end{figure}

For the standard $\Lambda$CDM model, two cases are investigated: a spatially
flat Universe, where~(\ref{luminositydistanceredshift}) is used to calculate
the luminosity distance and the general case of a curved Universe,
where~(\ref{luminositydistanceredshift_curv}) is employed. The best-fit
parameter values, the $1\sigma$ errors and the corresponding $\chi^2$ values
for the Riess07 (gold and gold+silver) and the WV07+HST datasets are listed in
Table~\ref{tb:SN_lcdm}. The confidence limits obtained in the $(\Omega_{\rm M},
\Omega_\Lambda)$ plane are shown in Fig.~\ref{fig:SN_BAO_Hz_LCDM_cont},
together with other astrophysical constraints which will be discussed in the
following sections. The fits to the three datasets are compatible and in
agreement with those presented elsewhere~\cite{HST,essence,davis}. In both
cases investigated
---for a spatially flat and for a curved Universe--- the SN data provide stringent
constraints on the cosmological parameters.

\begin{table}[ht]
\begin{center}
\begin{tabular}{| l | c c c | c c c |} \hline
 \T\BB & \multicolumn{3}{c|}{Flat Universe} & \multicolumn{3}{c|}{Curved Universe} \\ \hline
  Dataset\T\BB & $\Omega_{\rm M}$ & $\chi^2$ & $\chi^2/{\rm dof}$ & $(\Omega_{\rm M}, \Omega_{\Lambda})$ & $\chi^2$ & $\chi^2/{\rm dof}$ \\ \hline
  Riess07:gold\T      & $0.308\pm0.020$ & 160 & 0.88 & $(0.48, 0.95)$ & 156 & 0.87 \\
  Riess07:gold+silver & $0.295\pm0.017$ & 369 & 1.43 & $(0.55, 1.10)$ & 356 & 1.39 \\
  WV07+HST            & $0.259\pm0.019$ & 196 & 1.02 & $(0.33, 0.85)$ & 195 & 1.03 \\ \hline
\end{tabular}
\end{center}
\caption{\it Fits to the $\Lambda$CDM model parameter $\Omega_{\rm M}$,
assuming a flat Universe, and to parameters $\Omega_{\rm M}$ and
$\Omega_{\Lambda}$, when a curved one is considered. The values favoured by the
Riess07:gold, the Riess07:gold+silver and the WV07+HST datasets are listed,
together with the corresponding $\chi^2$ values.} \label{tb:SN_lcdm}
\end{table}

For the super-horizon model~\cite{riotto}, the results of the fits to the three
supernova datasets are presented in Table~\ref{tb:SN_lcdm}. The favoured
parameter values for these datasets are compatible and in full agreement with
those found in an earlier analysis \cite{EMMN}. The goodness of fit for this
model is comparable with the one for the standard $\Lambda$CDM model, as shown
by comparing the ratio of the $\chi^2$ to the number of degrees of freedom
(\emph{dof}).

\begin{table}[H]
\begin{center}
\begin{tabular}{| l | c c c |} \hline
  Dataset\T\BB        & $\Psi_{\ell0}$ & $\chi^2$ & $\chi^2/{\rm dof}$ \\ \hline
  Riess07:gold\T      & $-0.71\pm0.05$ &      167 & 0.92 \\
  Riess07:gold+silver & $-0.78\pm0.05$ &      383 & 1.48 \\
  WV07+HST            & $-0.90\pm0.07$ &      200 & 1.05 \\ \hline
\end{tabular}
\end{center}
\caption{\it Fits to the super-horizon model parameter $\Psi_{\ell0}$. The
values favoured by the Riess07:gold, the Riess07:gold+silver and the WV07+HST
datasets are listed, together with the corresponding $\chi^2$ values.}
\label{tb:SN_shcdm}
\end{table}

In the same manner, Q-cosmology models are compared against the recent
supernova data, including the off-shell effects, which are parametrised in the
Hubble rate~(\ref{eq:formulaforfit}). In general, this parametrisation of the
model is described by three parameters, $\delta$, $\Omega_\delta$ and
$\Omega_3$, which serve as free parameters of the fitting function to be
determined by the data analysis. The results for a simple case where the
parameter $\delta$ is kept fixed and equal to four are listed in
Table~\ref{tb:SN_qcosmo}. The value $\delta=4$ is chosen to match the results
of similar previous studies in Q-cosmology~\cite{EMMN} for the exotic scaling
of dark matter contributions, induced by dilatons, and dark radiation in brane
models without dilatons. The current analysis shows that the Q-cosmology model
fits equally well the supernova data as the $\Lambda$CDM model. The results for
the Riess07:gold and the WV07+HST datasets are consistent at the $1\sigma$
level. The Riess07:gold+silver dataset yields results compatible to the
Riess07:gold sample, but slightly different to the WV07+HST. Nevertheless, this
does not constitute a significant discrepancy, since the `silver' sample
comprises astrophysical objects whose classification as type-Ia supernovae is
uncertain \cite{HST}. The estimated parameters are compatible with those
derived in previous studies~\cite{EMMN}.

\begin{table}[H]
\begin{center}
\begin{tabular}{| l | c c c c c |} \hline
  Dataset\T\BB & $\Omega_3$ & $\Omega_4$ & $\Omega_2$ & $\chi^2$ & $\chi^2/{\rm dof}$ \\ \hline
  Riess07:gold\T      & $-3.4\pm0.8$ &  $1.2\pm0.3$  & $3.2\pm0.9$ & 157 & 0.87 \\
  Riess07:gold+silver & $-4.5\pm0.6$ & $1.61\pm0.25$ & $3.9\pm0.7$ & 357 & 1.39 \\
  WV07+HST          & $-2.8\pm0.5$ & $0.86\pm0.22$ & $2.9\pm0.5$ & 195 & 1.02 \\
  \hline
\end{tabular}
\end{center}
\caption{\it Fits to the Q-cosmology parameters $\Omega_3$ and $\Omega_4$ for a
fixed value $\delta=4$. The values favoured by the Riess07:gold, the
Riess07:gold+silver and the WV07+HST datasets are listed, together with the
corresponding $\chi^2$ values. $\Omega_2$ is determined by the other densities
so that $\Omega_4+\Omega_3+\Omega_2=1$. } \label{tb:SN_qcosmo}
\end{table}

\begin{figure}[htb]
\begin{center}
\includegraphics[width=0.45\linewidth]{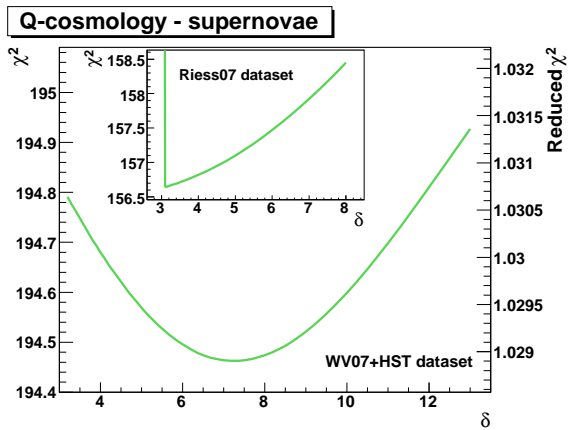}
\end{center}
\caption{\it Minimum $\chi^2$ and $\chi^2/{\rm dof}$ for the Q-cosmology model
and the WV07+HST SN dataset for various values of the $\delta$ parameter.
Inset: the same for the Riess07:gold dataset.} \label{fig:Qcosmo_SN_chi2}
\end{figure}

The analysis was subsequently performed treating $\delta$ as a free parameter.
In Fig.~\ref{fig:Qcosmo_SN_chi2}, the minimum $\chi^2$ achieved for fixed
values of $\delta$ leaving $\Omega_3$ and $\Omega_\delta$ unconstrained is
plotted versus the chosen $\delta$. Despite the fairly low $\chi^2$ value
achieved ($\chi^2/{\rm dof}=1.029$), the goodness of fit does not vary strongly
with $\delta$. For the WV07+HST dataset, the minimum value of $\chi^2$ is
$\sim194.5$, hence the 68\% confidence interval is defined by the $\chi^2<198$
condition, corresponding to a quite large range of favoured values for
$\delta$. This feature (the weak dependence of the fit results on $\delta$) is
also evident when the analysis is repeated for the Riess07:gold sample (cf.\
Fig.~\ref{fig:Qcosmo_SN_chi2}, inset), where the best-fit value for $\delta$
shifts from $\delta\simeq7$ (WV07+HST dataset) to $\delta\simeq3.1$. The
special case of $\delta=3$, yielding a $\chi^2$ of $\sim 181$ (Riess07:gold
dataset), is of no relevance to the Q-cosmology model, corresponding to a
generic matter-dominated Universe with no cosmological constant, already
disfavoured by astrophysical data. To recapitulate, Q-cosmology models fit the
supernovae data very well, with the analysis yielding a strong correlation
between the cosmological parameters rather than providing a concrete
determination of them. The latter may be achieved in conjunction with other
astrophysical data, as we shall discuss in Section~\ref{sc:discussion}, where
confidence intervals for this analysis will be presented.

\subsection{$H(z)$ constraints from distant galaxies}

\begin{figure}[htb]
\begin{center}
\includegraphics[width=0.45\linewidth]{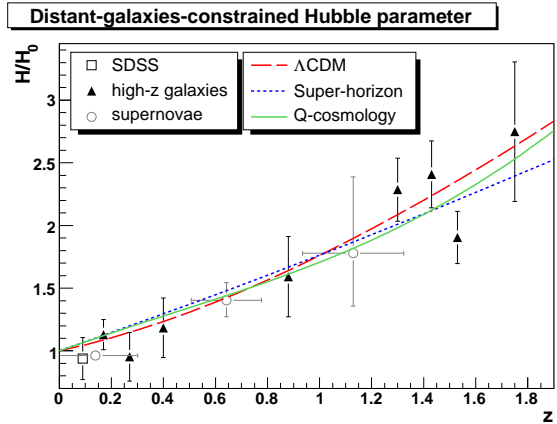}
\end{center}
\caption{\it Reduced Hubble parameter $H(z)/H_0$ as measured by SDSS
(square)~\cite{SDSS} and distant galaxies (triangles)~\cite{Hz} and for the
$H_0$ value measured by WMAP3~\cite{wmap}. The predictions for the following
cosmological models are drawn: (i) $\Lambda$CDM model for $(\Omega_{\rm M},
\Omega_\Lambda) = (0.2,0.7)$ (red long-dashed line); (ii) super-horizon model
for $\Psi_{\ell0} = -0.56$ (blue short-dashed line); and (iii) Q-cosmology
model for $\delta=4$, $\Omega_3=-1$ and $\Omega_\delta=0.25$ (green thick solid
line). The measurements from supernovae data~\cite{HST} are also shown
(circles). } \label{fig:Hubble_fit}
\end{figure}

We now attempt to further constrain the cosmological models under study by
performing fits to the Hubble parameter dependence on redshift, $H(z)$, as
determined~\cite{Hz,SDSS} using measurements of differential ages of passively
evolving galaxies following the method described in Ref.~\cite{Hz_method}. This
method was applied on a sample of distant galaxies consisted by the Gemini Deep
Deep Survey (GDDS)~\cite{GDDS} and archival data~\cite{galaxies}, spanning a
redshift range of $0<z<2$. The ratio $H(z)/H_0$ of these Hubble-rate
measurements to the Hubble constant $H_0 = (73.5\pm3.2)~{\rm
Mpc^{-1}\,km\,s^{-1}}$, measured by WMAP3~\cite{wmap}, is shown in
Fig.~\ref{fig:Hubble_fit}. The point at $z\sim0.1$ was derived~\cite{SDSS} by
the Luminous Red Galaxy (LRG) sample reported in the Sloan Digital Sky Survey
(SDSS)~\cite{SDSS_LRG}. For comparison, the points derived by the latest
supernovae data released by the HST~\cite{HST} are also plotted, even through
they were not taken into account in the fits discussed in the following.

Here, the Hubble parameter as a function of the redshift, given in
Eqs.~(\ref{eq:formulaforfit}),~(\ref{eq:lcdm}) and~(\ref{eq:shcdm}) for the
three cosmological models, is directly fitted to the aforementioned data
points. The model predictions for the best-fit parameter values are given in
Fig.~\ref{fig:Hubble_fit} for the three models.

\begin{figure}[htb]
\begin{center}
\includegraphics[width=0.45\linewidth]{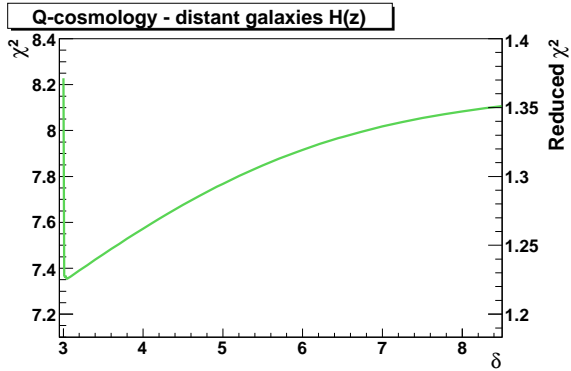}
\end{center}
\caption{\it Minimum $\chi^2$ and $\chi^2/{\rm dof}$ for the Q-cosmology model
and the galactic $H(z)$ determination for various values of the $\delta$
parameter.} \label{fig:Qcosmo_Hz_chi2}
\end{figure}

The $\Lambda$CDM model fits the data very well: if a spatially flat Universe is
assumed, a value of $\Omega_{\rm M}=0.27\pm0.04$ is derived with $\chi^2=6.47$
($\chi^2/{\rm dof}=0.809$), while in the general case of a curved Universe, the
best-fit values are $\Omega_{\rm M}=0.2\pm0.3$ and $\Omega_{\Lambda} =
0.7\pm0.6$ with $\chi^2=6.46$ ($\chi^2/{\rm dof}=0.923$). The $H(z)$ curve for
the latter parameters is shown in Fig.~\ref{fig:Qcosmo_Hz_chi2} and the fit
result with the corresponding $1\sigma$-error is superimposed on the SN results
in Fig.~\ref{fig:SN_BAO_Hz_LCDM_cont}. It is evident that this measurement
provides a quite loose constraint on the $\Lambda$CDM model with errors of the
order of 15\% (100\%) for a flat (curved) Universe and it merely confirms the
supernova analysis results.

On the other hand, the $H(z)$ measurement provides a stringent constraint on
the super-horizon model parameter: $\Psi_{\ell0}=-0.56\pm0.09$ with a fairly
good fit of $\chi^2=6.98$ ($\chi^2/{\rm dof}=0.872$). The respective $H(z)$
curve is drawn in Fig.~\ref{fig:Qcosmo_Hz_chi2}. This $\Psi_{\ell0}$ value,
however, is more than $2\sigma$ higher than the one favoured by the
high-redshift supernovae of the WV07+HST dataset, presented in
Table~\ref{tb:SN_shcdm}. An elaborate discussion on this model follows in the
next section and in Sec.~\ref{sc:discussion}, in view of the baryonic
oscillations data.

For the Q-cosmology model, the same procedure as for the supernovae is
followed. For a fixed value of $\delta=4$, the best-fit parameters are
$\Omega_3=0.2\pm0.3$ and $\Omega_\delta=-1.0\pm1.0$ with $\chi^2=7.57$
($\chi^2/{\rm dof}=1.08$). The fit is not as good as for the other two models,
nevertheless it is still acceptable with a ratio of $\chi^2$ over number of
degrees of freedom near unity (see also the Q-cosmology prediction in
Fig.~\ref{fig:Hubble_fit}). The errors on the parameters are large, but
---as will shall see in Sec.~\ref{sc:discussion}--- they are correlated and
the parameters are eventually sufficiently constrained especially when combined
with other observational sources.

Likewise, the analysis was repeated treating $\delta$ as a free parameter. The
evolution of the minimum $\chi^2$ obtained for various values of $\delta$ (and
the corresponding best-fit values for $\Omega_3$ and $\Omega_\delta$) is shown
in Fig.~\ref{fig:Qcosmo_Hz_chi2}. The global minimum value of $\chi^2=7.35$
($\chi^2/{\rm dof}=1.2$) occurs at $\delta\simeq3$, however as for the
supernovae, the quality of fit does not change dramatically with $\delta$, the
68\% confidence level being at $\chi^2=10.9$. This outcome is further discussed
in Sec.~\ref{sc:discussion} in conjunction with other astrophysical data.

The fits to all three models were likewise repeated for the same set of $H(z)$
points with the addition of the three points measured by HST. The derived model
parameters and the goodness of fit were not significantly different than the
ones reported above, since the HST points and their errors lie along the line
connecting the galactic measurements.

\subsection{Baryon acoustic oscillations}

Complementary information on cosmological models is provided by data on Baryon
Acoustic Oscillations (BAO), which show up in the patterned distribution of
galaxies on very large scales ($\gtrsim100~{\rm Mpc}$). The SDSS~\cite{baryon}
survey covers the region $0.16<z<0.47$, i.e.\ the typical redshift is $z_{\rm
BAO}=0.35$. Among other observables, the SDSS collaboration has measured the
dilation scale, $D_V$, defined as~\cite{baryon}:
\begin{equation}\label{eq:Dv}
D_V(z) = \left[ D_M^2(z) \frac{cz}{H(z)}\right]^{1/3},
\end{equation}
where $D_M$ is the co-moving angular diameter distance at redshift $z$.
Equation~(\ref{eq:Dv}) presupposes a single-scale approximation in the
redshift-to-distance conversion and an equivalent treatment between the
line-of-sight and the transverse dilation. These assumptions are valid for the
fiducial $\Lambda$CDM model for $(\Omega_{\rm M},
\Omega_\Lambda)\simeq(0.3,0.7)$ for redshifts around 0.35~\cite{baryon}. They
equally apply to the Q-cosmology model under discussion, as the Hubble
parameter at $z\simeq0.35$ does not deviate substantially from the fiducial
$\Lambda$CDM, as shown in Fig.~\ref{fig:Hubble_fit}. The measured value of the
dilation scale at $z_{\rm BAO}=0.35$ is $D_V(0.35)=1370\pm64~{\rm
Mpc}$~\cite{baryon}.

Additionally, the observable $\Omega_{\rm M}h^2$, where $h$ is the reduced
Hubble constant, has been measured by SDSS: $\Omega_{\rm M}h^2=0.130\pm0.010$.
Cosmological models predictions are consequently performed by the SDSS
collaboration~\cite{baryon} against the observable $A$, defined as:
\begin{equation}\label{eq:Adefined}
A \equiv D_V(z_{\rm BAO}) \frac{\sqrt{\Omega_{\rm M}H_0^2}}{c\,z_{\rm BAO}}
\end{equation}
with a measured value of $A=0.469\pm0.017$. For a constant equation of state of
dark energy, the quantity $A$ is expressed as~\cite{Guo}:
\begin{multline}
A = \frac{\sqrt{\Omega_{\rm M}}}{z_{\rm BAO}} \left[ \frac{z_{\rm
BAO}}{|\Omega_k|E(z_{\rm BAO})}\right]^{1/3} \times \begin{cases}%
 \sin^{2/3}\left( \sqrt{|\Omega_k|}\int_0^{z_{\rm BAO}} \frac{dz}{E(z)}\right), & \mbox{for a closed Universe,} \\
 \left[ \sqrt{|\Omega_k|}\int_0^{z_{\rm BAO}} \frac{dz}{E(z)}\right]^{2/3}, & \mbox{for a flat Universe,} \\
 \sinh^{2/3}\left( \sqrt{|\Omega_k|}\int_0^{z_{\rm BAO}} \frac{dz}{E(z)}\right), & \mbox{for an open Universe.}
\end{cases}\label{eq:Acurved}
\end{multline}
where $E(z)\equiv H(z)/H_0$ is the reduced Hubble parameter.

The impact of the measurement of observable $A$ on the $\Lambda$CDM model is
depicted at $1\sigma$ level on top of other constraints in
Fig.~\ref{fig:SN_BAO_Hz_LCDM_cont}. As shown also in Ref.~\cite{Guo}, BAO
impose a further constraint (besides the observations of type-Ia supernovae) on
the $\Lambda$CDM model parameters, $\Omega_{\rm M}$ and $\Omega_\Lambda$. If a
spatially flat geometry for the Universe is assumed, the matter density is
determined by observable $A$ to be $\Omega_{\rm
M}=0.273\pm0.025$~\cite{baryon}.

In the Q-cosmology case, as mentioned in Sec.~\ref{sc:qcosmo}, the ordinary and
dark matter component appears in the Hubble parameter~(\ref{eq:formulaforfit})
mixed with exotic contributions. Therefore, the measurement of $\Omega_{\rm
M}h^2$ ---and of the observable $A$ consequently--- does not constrain the
Q-cosmology model parameters $\Omega_3$ and $\Omega_\delta$. Nonetheless, it
does set limits to the sum of the matter constituents of present-day densities
$\Omega_3$ and $\Omega_\delta$. The predictions of Q-cosmology are hence
compared against another observable, $B$, involving the dilation scale $D_V$
reduced so as to remove the dependence on the current value of the Hubble
constant, $H_0$:
\begin{equation}\label{eq:B_definition}
B \equiv \frac{H_0}{c}D_V(z_{\rm BAO}) = \left[ \left( \int_0^{z_{\rm
BAO}}\frac{dz}{E(z)} \right)^2 \frac{z_{\rm BAO}}{E(z_{\rm BAO})} \right]^{1/3}
\end{equation}
For a Hubble constant $H_0 = (73.5\pm3.2)~{\rm Mpc^{-1}\,km\,s^{-1}}$, as
measured by WMAP3~\cite{wmap} ---also employed in the distant galaxies
analysis---, we derive a measured value of $B=0.334\pm0.021$. Although, the BAO
observations may not provide a determination of the Q-cosmology parameters,
nevertheless, as we shall see in Sec.~\ref{sc:discussion}, they constrain such
models when combined with other astrophysical data.

\begin{figure}[htb]
\begin{center}
\includegraphics[width=0.45\linewidth]{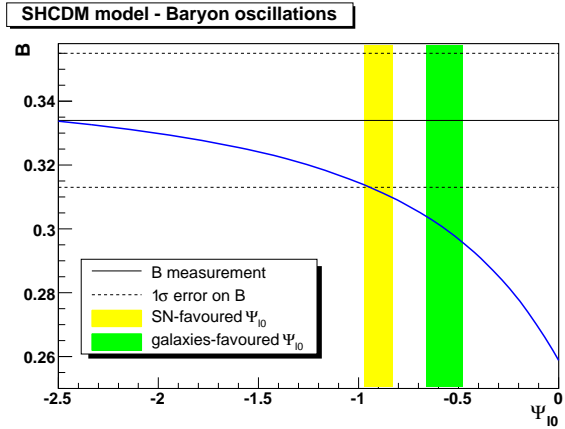}
\end{center}
\caption{\it Observable $B$ prediction versus $\Psi_{\ell0}$ for the
super-horizon model. The horizonal lines show the measured value of $B$
($\pm1\sigma$); the left-hand (right-hand) vertical band shows the value and
$\pm1\sigma$ error favoured by the supernovae (distant galaxies) data. }
\label{fig:BAO_B_SHCDM}
\end{figure}

As far as the super-horizon model is concerned, the independent measurement of
$\Omega_{\rm M}h^2$~\cite{baryon} by BAO, implying a matter density
$\Omega_{\rm M}\simeq0.27$, practically excludes it, since it predicts a matter
density equal to unity~\cite{riotto}. Yet the model parameter range allowed by
observable $B$ (devoid of any dependence on $\Omega_{\rm M}$), shown in
Fig.~\ref{fig:BAO_B_SHCDM} (blue curve), marginally agrees with the
supernovae-favoured value for $\Psi_{\ell 0}$.

\section{Discussion and interpretation}\label{sc:discussion}

For the conventional $\Lambda$CDM model, where a redshift-independent equation
of state $w_{\rm DE}=-1$ for the dark energy component is assumed, the
constraints imposed by supernovae, distant galaxies and baryonic oscillations
are reviewed in Table~\ref{tb:lcdm} and in Fig.~\ref{fig:SN_BAO_Hz_LCDM_cont}
as well. The determination of the Hubble parameter by high-redshift galaxies
loosely constrain the model parameters. The other two astrophysical
observations, on the other hand, set a combined parameter determination around
$(\Omega_{\rm M}, \Omega_{\Lambda})\simeq(0.28,0.75)$, as shown in
Fig.~\ref{fig:SN_BAO_Hz_LCDM_cont}.

\begin{table}[H]
\begin{center}
\begin{tabular}{| l | c c c | c c c |} \hline
 \T\BB & \multicolumn{3}{c|}{Flat Universe} & \multicolumn{3}{c|}{Curved Universe} \\ \hline
  Data \T\BB & $\Omega_{\rm M}$ & $\chi^2$ & $\chi^2/{\rm dof}$ & $(\Omega_{\rm M}, \Omega_{\Lambda})$ & $\chi^2$ & $\chi^2/{\rm dof}$ \\ \hline
  Supernovae (WV07+HST) \T\BB & $0.259\pm0.019$ & 196 & 1.02 & $(0.33,0.85)$ & 195 & 1.03 \\
  Distant galaxies ($H(z)$) & $0.27\pm0.04$   &6.47 &0.809 & $(0.2,0.7)$ & 6.46 & 0.923 \\
  Baryon oscillations ($A$)                  & $0.273\pm0.025$ & -- & -- & \multicolumn{3}{c|}{\it see Fig.~\ref{fig:SN_BAO_Hz_LCDM_cont}} \\
  \hline
\end{tabular}
\end{center}
\caption{\it $\Lambda$CDM model parameter $\Omega_{\rm M}$, assuming a flat
Universe, and parameters $\Omega_{\rm M}$ and $\Omega_{\Lambda}$ for a curved
one resulting from type-Ia supernovae (WV07+HST), distant galaxies ($H(z)$) and
baryon acoustic oscillations ($A$).} \label{tb:lcdm}
\end{table}

The results for the super-Hubble model, which is based on cosmological
perturbations without introducing any dark energy component, are summarised in
Table~\ref{tb:shcdm} and Fig.~\ref{fig:BAO_B_SHCDM}. The type-Ia supernovae
data and the galactic $H(z)$ determination set incompatible constraints at
$2\sigma$ level on the model sole parameter, $\Psi_{\ell0}$. Furthermore,
baryon oscillation observations lead to an independent measurement of
$\Omega_{\rm M}\simeq0.27$, as opposed to the model-predicted unity value for
the matter density. The $\Psi_{\ell0}$ values which are consistent with the
observable $B$ measurement, on the other hand, are barely in agreement with the
supernova fit result. In conclusion, the super-horizon model appears
over-constrained when confronted by different sources of astrophysical
observations and it is excluded by this analysis.

\begin{table}[H]
\begin{center}
\begin{tabular}{| l | c c c |} \hline
  Data \T\BB & $\Psi_{\ell0}$ & $\chi^2$ & $\chi^2/{\rm dof}$ \\ \hline
  Supernovae (WV07+HST) \T\BB & $-0.90\pm0.07$ & 200  & 1.05  \\
  Distant galaxies ($H(z)$)  & $-0.56\pm0.09$ & 6.98 & 0.872 \\
  Baryon oscillations ($B$)   & $<-0.8$        & --   & --  \\
  \hline
\end{tabular}
\end{center}
\caption{\it Super-horizon model parameter $\Psi_{\ell0}$, rendered by type-Ia
supernovae (WV07+HST), distant galaxies ($H(z)$) and baryon acoustic
oscillations ($B$).} \label{tb:shcdm}
\end{table}

\begin{figure}[htb]
\begin{center}
\includegraphics[width=0.45\linewidth]{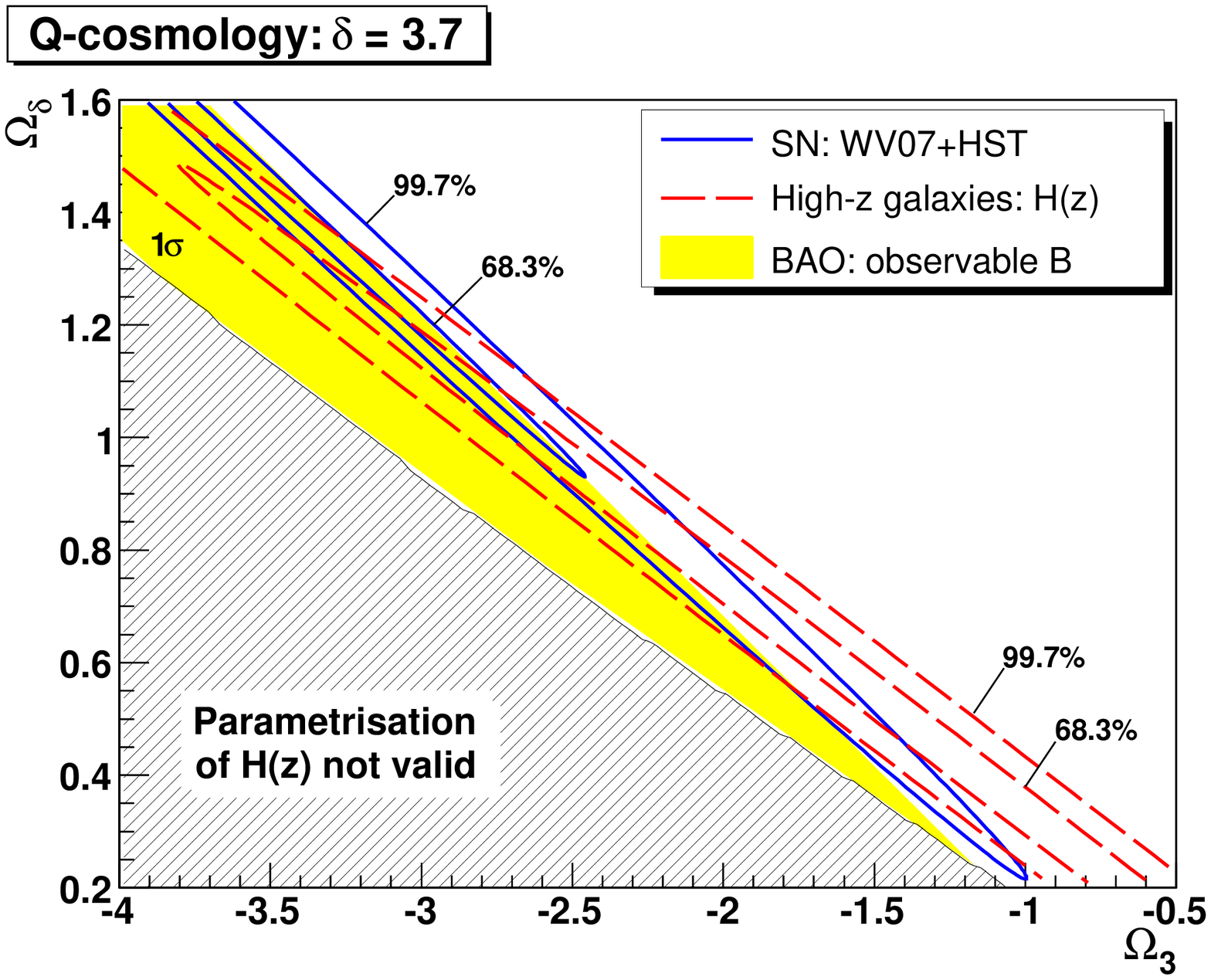}
\includegraphics[width=0.45\linewidth]{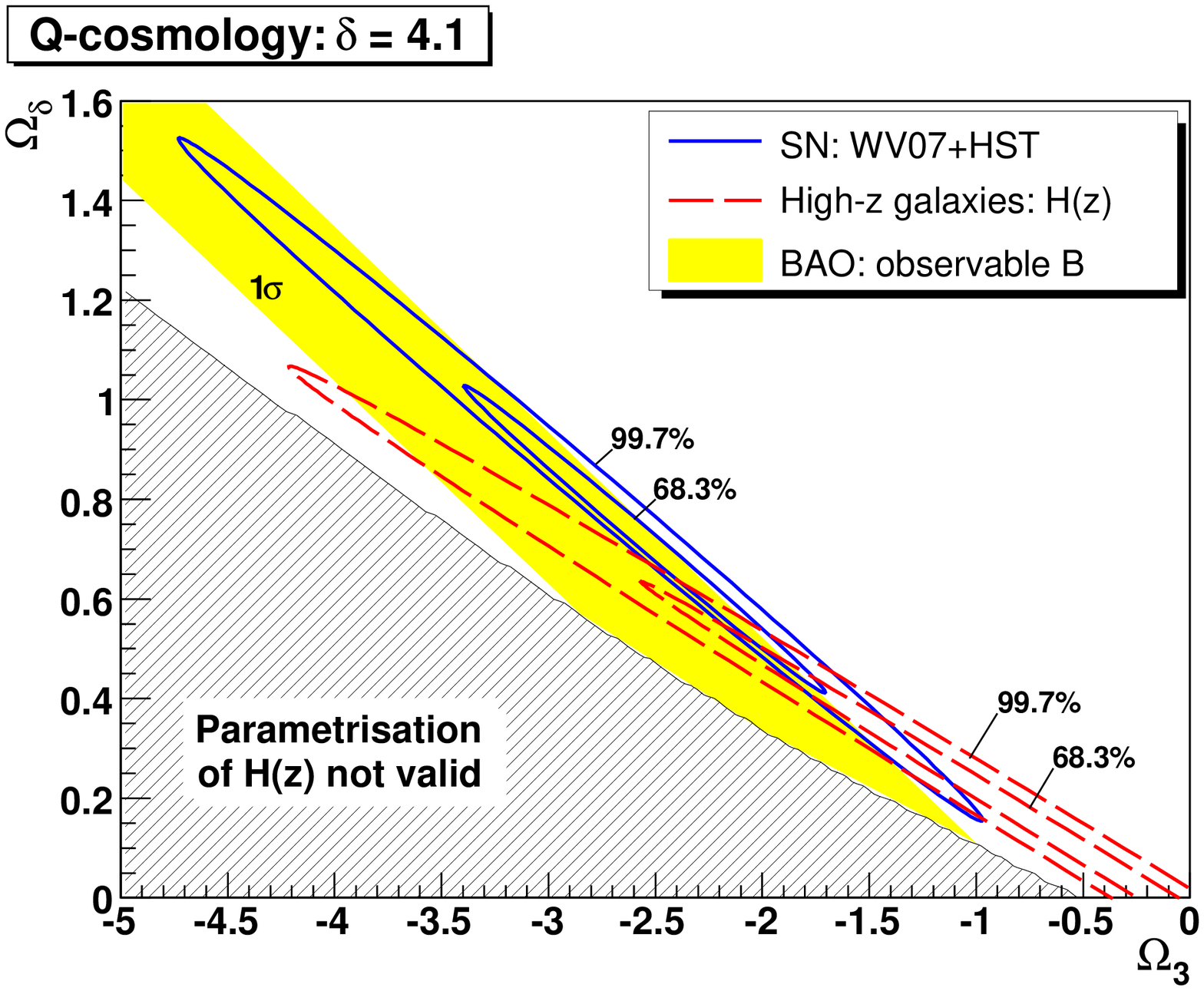}
\includegraphics[width=0.45\linewidth]{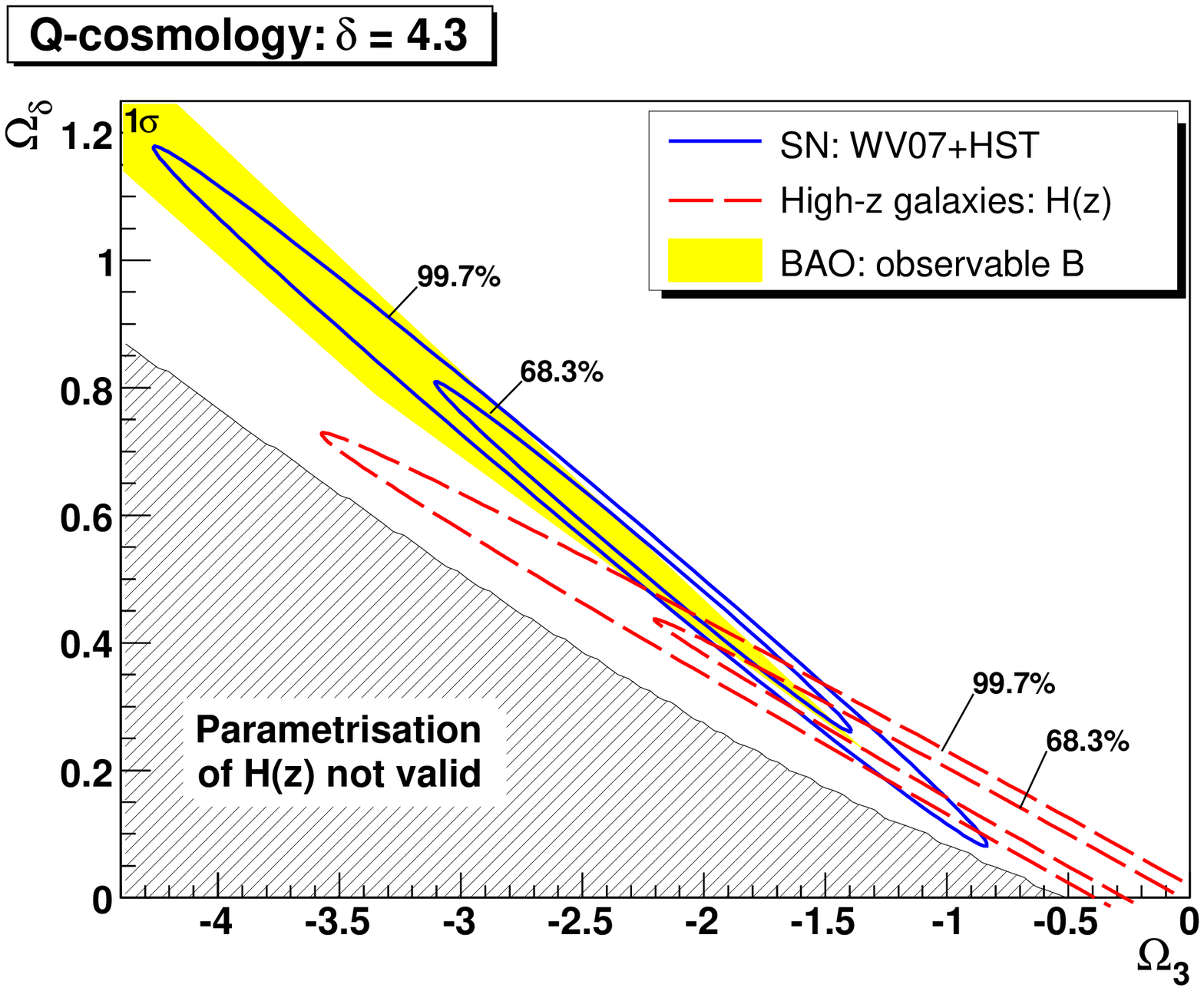}
\end{center}
\caption{\it Observational constraints on Q-cosmology for $\delta =
3.7,\:4.1,\:4.3$ from top to bottom, respectively: (a) 68.3\% and 99.7\% C.L.\
contours for the WV07+HST supernova dataset (blue solid line); (b) 68.3\% and
99.7\% C.L.\ contours for the galactic $H(z)$ (red dashed line); and (c)
$1\sigma$ region from the observable $B$ of baryon acoustic oscillations
(yellow area).} \label{fig:Qcosmo_all}
\end{figure}

We now proceed on discussing the (off-shell) Q-cosmology models. The allowed
parameter regions are shown in Fig.~\ref{fig:Qcosmo_all} for various values of
the parameter $\delta$ related to the exotic scaling of dark matter: $\delta =
3.7,\:4.1,\:4.3$. All three astrophysical constraints are presented: the
WV07+HST supernovae (68.3\% and 99.7\% C.L.), the distant galaxies (68.3\% and
99.7\% C.L.) and the observable $B$ of BAO ($1\sigma$). The $(\Omega_3,
\Omega_\delta)$ region for which the parametrisation~(\ref{eq:formulaforfit})
for the Hubble parameter is not valid (i.e.\ $H^2(z)<0$) for $0<z<2$ is also
indicated. The constraints from supernova data largely overlap and lie in
parallel with those from baryon acoustic oscillations. This stems from the fact
that the corresponding observables
---luminosity distance $d_L$ and dilation scale $D_V\propto B$--- are both
sensitive to $\int_0^z\frac{dz'}{H(z')}$. However, the typical redshifts $z$
for these measurements are quite different: $z_{\rm SN}=2$ for the supernovae
as opposed to $z_{\rm BAO}=0.35$ for BAO, rendering independent bounds.

\begin{figure}[htb]
\begin{center}
\includegraphics[width=0.45\linewidth]{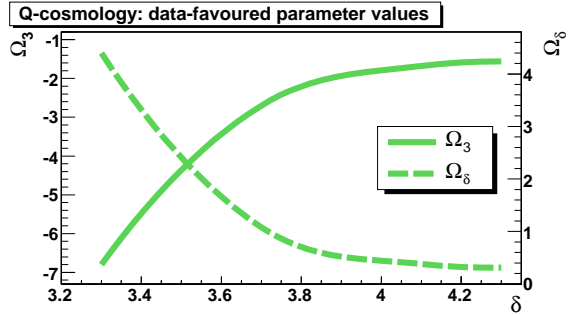}
\end{center}
\caption{\it Astrophysical data-favoured values of Q-cosmology parameters as
defined from the overlap of the $1\sigma$ regions from supernovae, distant
galaxies and baryon oscillations data.} \label{fig:Qcosmo_overlap}
\end{figure}

The parameter $\delta$ may be restricted by considering the overlap of all three
cosmological constraints. For values $\delta\lesssim3.3$, the 68\% confidence
intervals for BAO and SN do not overlap. Furthermore, the $1\sigma$ contours
for the SN and the galaxies partially coincide only for $\delta\lesssim4.3$.
Hence, the allowed range for $\delta$ is fairly wide:
$3.3\lesssim\delta\lesssim4.3$. The approximate values for the other two
parameters, $\Omega_3$ and $\Omega_\delta$, as defined by the overlapping
regions of the $1\sigma$ intervals for all three astrophysical data are given
in Fig.~\ref{fig:Qcosmo_overlap} as a function of $\delta$. These values
represent the trend of the densities $\Omega_i$ with varying $\delta$ rather
than a precise determination of model parameters (hence the thick curves). To
recapitulate, the allowed ranges for the Q-cosmology parameters defined in the
parametrisation~(\ref{eq:formulaforfit}), as restricted by current type-Ia
supernovae, high-redshift galaxies and baryon acoustic oscillations are:
$3.3\lesssim\delta\lesssim4.3$, $-7\lesssim\Omega_3\lesssim-1.5$ and
$0.2\lesssim\Omega_\delta\lesssim4.5$ with the correlations defined in
Fig.~\ref{fig:Qcosmo_overlap}.

The negative-energy dust contribution appears essential in order to provide
fits to the astrophysical data, and, as we have stressed previously, it may not
be inconsistent with positive-energy theorems of the underlying microscopic
cosmological model. In brane-world scenarios, for instance, such
negative-energy dust might represent effective four-dimensional contributions
to the energy density of KK-compactification (massive) graviton modes, as
measured by an observer on the brane, while the corresponding bulk
energy-density contributions are perfectly positive definite. In Q-cosmology
models, such negative energies might also be due to the coupling with the non
constant dilaton-quintessence field, including higher-string-loop terms, again
without implying a conflict with positive-energy conditions. In addition to
this negative dust energy, the data also provide evidence for (positive)
exotic-scaling contributions to the energy density of the form $a^{-\delta}$.
The value of $\delta = 4$ is included in the phenomenologically allowed range.
Such terms may therefore be interpreted as either dark radiation, in the
context of brane models~\cite{brane}, or exotic dark-matter contributions in
Q-cosmologies~\cite{emn,lmn,diamandis2}, due to the coupling with non-trivial
quintessence dilatons, or even {\it both} types in the case of brany
non-critical-string dilaton-quintessence models~\cite{emnw}.

Stringent constraints on these brane and off-shell Q-cosmological models may be
set by the WMAP measurements~\cite{wmap} of the cosmic microwave background
(CMB), especially the position of the (acoustic) peaks in the spectrum.
Nevertheless, since CMB provides evidence on the last scattering surface at a
redshift of $z\sim1100$, more phenomenological input is required in order to
embark on the pertinent analysis. Reversely, astrophysical data analysis
provides a handle to exclude regions in parameter space and to guide the
phenomenological studies.

\section{Conclusions and outlook}\label{sc:conclu}

In this paper, we have performed an analysis on data collected from three
distinct astrophysical sources, namely type-Ia supernovae ($z\lesssim2$),
distant galaxies ($z\lesssim2$) ---via the determination of the Hubble
expansion rate--- and baryon acoustic oscillations observed in luminous red
galaxies ($0.1\lesssim z\lesssim0.5$). We compared them with the respective
predictions of the conventional $\Lambda$CDM model~\cite{carroll}, the
super-Hubble model~\cite{riotto} and the
Q-cosmology~\cite{emnw,diamandis,diamandis2,lmn,EMMN}. The $\Lambda$CDM model,
as expected, fits the data very well and is mainly constrained by the supernova
and BAO data. The super-horizon model, on the other hand, is excluded by the
BAO-measured matter density and yields contradictory (at $2\sigma$ level)
parameter determinations by SN and galactic data.

Finally, the spatially-flat Q-cosmology predictions, as parametrised
in~(\ref{eq:formulaforfit}) for $z\lesssim2$, fits the data very well providing
an alternative scenario to account for the dark energy component of the
Universe. The three model parameters are not univocally determined; their
allowed range and the correlation between them is rather defined. The parameter
$\delta$, associated to the exotic scaling of dark matter in the context of
Q-cosmology, is restricted in the range $3.3\lesssim\delta\lesssim4.3$, whilst
the present values of (exotic) matter densities, $\Omega_3$ and
$\Omega_\delta$, vary with $\delta$ as shown in Fig.~\ref{fig:Qcosmo_overlap}.

The fact that the value of $\delta=4$ is included in the allowed range, also
points towards dark radiation terms in brane models~\cite{brane}. The data also
point towards negative-energy dust contributions, which could be due to either
dark-energy dilaton terms in Q-cosmologies~\cite{EMMN,emnw}, or Kaluza-Klein
compactification (massive) graviton modes in brane-inspired
modes~\cite{kaluza}.

Further detailed studies, such as the theoretical determination of the position
of the acoustic peaks in the CMB spectrum, using the underlying formalism of
the above models, or the measurement of the pertinent (complicated, in general
$z$-dependent) equation of state, as, for instance, in the analysis of
Refs.~\cite{Hz,wz}, are certainly required in order to provide further evidence
that might discriminate among the various models. The important point that our
analysis, however, brought hopefully forward, is the fact that alternative,
unconventional Cosmologies are certainly compatible with the current data, and
appear on an equal footing to the standard $\Lambda $CDM cosmological model.
Thus, surprises on the possibility of discovering completely new and
unconventional physical phenomena in the foreseeable future cannot be excluded.

\section*{Acknowledgements}

We wish to thank John Ellis for discussions. The work of N.E.M.\ is partially
supported by the European Union through the FP6 Marie Curie Research and
Training Network \emph{UniverseNet} (MRTN-CT-2006-035863).

\end{document}